\documentclass[aip,apl,reprint,twocolumn]{revtex4-1}

\usepackage{graphicx}
\usepackage{dcolumn}
\usepackage{bm}
\usepackage{xfrac}
\usepackage{SIunits}
\usepackage{color}
\usepackage[usenames,dvipsnames]{xcolor}
\usepackage{hyperref}
\usepackage{footmisc}
\usepackage{fix-cm}
\usepackage{amsmath}

\usepackage[utf8]{inputenc}
\usepackage[T1]{fontenc}
\usepackage{mathptmx}

\newcommand{\ns}{n_{\mathrm{s}}}
\newcommand{\wc}{\omega_{\mathrm{res}}}
\newcommand{\QL}{Q_{\mathrm{L}}}

\begin{document}

\title{Tuning high-Q superconducting resonators by magnetic field reorientation}

\author{Christoph W. Zollitsch}
\email{c.zollitsch@ucl.ac.uk}
\affiliation{London Centre for Nanotechnology, University College London, 17-19 Gordon Street, London, WCH1 0AH, UK}

\author{James O'Sullivan}
\affiliation{London Centre for Nanotechnology, University College London, 17-19 Gordon Street, London, WCH1 0AH, UK}

\author{Oscar Kennedy}
\affiliation{London Centre for Nanotechnology, University College London, 17-19 Gordon Street, London, WCH1 0AH, UK}

\author{Gavin Dold}
\affiliation{London Centre for Nanotechnology, University College London, 17-19 Gordon Street, London, WCH1 0AH, UK}

\author{John J. L. Morton}
\affiliation{London Centre for Nanotechnology, University College London, 17-19 Gordon Street, London, WCH1 0AH, UK}
\affiliation{Dept.\ of Electronic \& Electrical Engineering, UCL, London WC1E 7JE, United Kingdom}

\date{\today}

\begin{abstract}
Superconducting resonators interfaced with paramagnetic spin ensembles are used to increase the sensitivity of electron spin resonance experiments and are key elements of microwave quantum memories. Certain spin systems that are promising for such quantum memories possess `sweet spots' at particular combinations of magnetic fields and frequencies, where spin coherence times or linewidths become particularly favorable. In order to be able to couple high-Q superconducting resonators to such specific spin transitions, it is necessary to be able to tune the resonator frequency under a constant magnetic field amplitude.
Here, we demonstrate a high-quality, magnetic field resilient superconducting resonator, using a 3D vector magnet to continuously tune its resonance frequency by adjusting the orientation of the magnetic field. The resonator maintains a quality factor of $> 10^5$ up to magnetic fields of $2.6\,\tesla$, applied predominantly in the plane of the superconductor. We achieve a continuous tuning of up to $30\,\mega\hertz$ by rotating the magnetic field vector, introducing a component of $5\,\milli\tesla$ perpendicular to the superconductor.

\end{abstract}



\maketitle

Superconducting co-planar microwave resonators allow for a variety of compact designs in conjunction with high-quality factors, and find applications in the sensitive readout of individual quantum systems and small ensembles \cite{Wallraff2004, Clarke2008, Fragner2008, Graaf2013, Bienfait2016, Bienfait2017, Probst2017} and the coupling of distinct physical systems \cite{Clarke2008, Mi2017, Samkharadze2018}. Superconducting resonators inductively coupled to atomic impurity spins form the basis of proposals for spin-based quantum memories \cite{Wesenberg2009, Wu2010, Grezes2014, Grezes2015, Morton2018} and have led to substantial advances in the detection limit of electron spin resonance \cite{Bienfait2016, Bienfait2017, Probst2017}.

The study of spins coupled to superconducting microwave resonators typically requires static magnetic fields in the range of several $100\,\milli\tesla$ to tune the spin Zeeman energy into resonance with the resonator. Superconducting resonators often exhibit limits in the quality factor ($< 10^5$) under the influence of such static magnetic fields \cite{Bothner2011, Graaf2012, Kwon2018}, and while previous studies have shown enhanced magnetic field resilience of high-quality factor ($> 10^5$) \cite{Samkharadze2016, Kroll2019}, these resonator designs were not optimized for high sensitivity spin sensing.
Furthermore, of particular interest in the context of long-lived spin-based quantum memories, are specific spin transitions which show an increased resilience to dominant sources of noise (e.g. magnetic or electric field noise) \cite{Wolfowicz2013, Morse2017, Ortu2018}. Prominent examples of systems with such magnetic field noise resilient transitions include bismuth donors in silicon, where the donor electron spin coherence time reaches seconds \cite{Wolfowicz2013}, as well as rare-earth dopants (e.g. Nd, Er or Yb) in Y$_2$SiO$_5$ \cite{Dold2019} reaching electron spin coherence times of $1\,\milli\second$ \cite{Ortu2018}. In the latter case, the additional presence of robust optical transitions leads to potential applications for microwave-to-optical quantum transducers. Common to all these applications is an optimum working point which is dictated by the spin species and sets both the magnetic field magnitude and the required resonator frequency at this given magnetic field. Matching the resonator frequency to the relevant spin transition is challenging due to fabrication uncertainties relating to film deposition and device patterning which affect frequency reproducibility – indeed this is becoming a wide-spread challenge in the field of kinetic inductance detectors \cite{McHugh2012} and quantum circuits \cite{Chen2012}. This challenge is further compounded by the additional frequency down-shift of the resonator due to an applied in-plane magnetic field, which needs to be accounted for before fabrication.
\textit{In-situ} frequency tunable resonators offer a practical route to adjust the resonator frequency, which increases the tolerance of fabrication uncertainties, and additionally offer the ability to study a spin system across a (small) frequency range. Several methods have been demonstrated for frequency-tuning superconducting resonators, including i) current-biasing through the signal line \cite{Asfaw2017, Adamyan2016}; ii) embedding SQUIDs into the resonator as magnetic-field tunable inductors \cite{Palacios2008, Sandberg2008, Kennedy2019}; and iii) simply applying global magnetic fields to tune the resonator frequency \cite{Healey2008, Samkharadze2016, Xu2019}. None of these approaches is ideally suited to the task of achieving strong coupling to noise-resilient spin transitions: they display a magnetic field resilience which is either limited \cite{Kennedy2019, Xu2019} or not investigated \cite{Palacios2008, Sandberg2008}, possess relatively low quality factors \cite{Asfaw2017} or rely on changing the overall magnetic field strength \cite{Adamyan2016, Samkharadze2016, Healey2008, Xu2019} (despite this value being determined by the chosen spin transition).

In this Article, we present a superconducting thin-film lumped element resonator (LER) tailored for a high resilience to static in-plane magnetic fields (up to $2.6\,\tesla$), and show how its frequency may be tuned by introducing an additional magnetic field component, perpendicular to the superconducting thin-film. In this way, we demonstrate frequency tunability of up to $30\,\mega\hertz$ (arising for a perpendicular magnetic field component of $5\,\milli\tesla$) while maintaining high-quality factors ($\QL > 10^5$). 

The resonator frequency $\omega_{\mathrm{res}} = 1/\sqrt{LC}$, where $L$ and $C$ are respectively the inductance and capacitance of the resonator \cite{Pozar1998}. The inductance can be further divided as $L = L_{\mathrm{G}} + L_{\mathrm{kin}}$, where $L_{\mathrm{G}}$ is the geometric inductance and $L_{\mathrm{kin}}$ is the kinetic inductance \cite{Goeppl2008}, arising from the finite inertia of the charge carriers \cite{Tinkham1975}, whose resulting effect is similar to an electromotive force on a charge in an inductor. To tune the resonator frequency, we exploit the dependence of $L_{\mathrm{kin}}$ on the Cooper pair density $\ns$, which takes the form $L_{\mathrm{kin}} \propto 1/n_{\mathrm{s}}$ \cite{Wallace1991, Watanabe1994}. Applying a static magnetic field reduces $\ns$, thus tuning the resonator to lower frequencies, and as long as the applied field does not exceed the first critical field, hysteretic effects in frequency tuning can be avoided \cite{Bothner2012}.  

\begin{figure}[t]
 \includegraphics[]{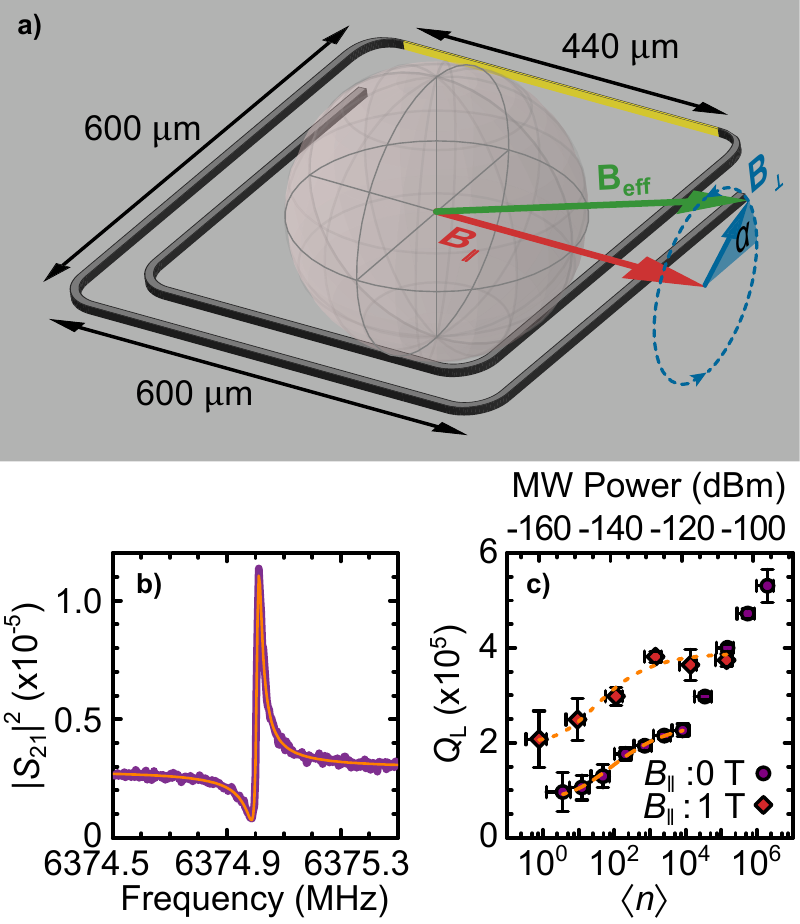}
 \caption{(a) Schematic of the lumped element NbN thin-film resonator, and the applied magnetic field $B_{\mathrm{eff}}$ with two components: $B_{\parallel}$ lies precisely in the plane of the superconducting film and nominally along the inductor wire (highlighted in yellow); while $B_{\perp}$ is defined perpendicular to $B_{\parallel}$ with an angle $\alpha$ to the plane of superconducting film. To tune the resonator frequency $\alpha$ is varied, while maintaining constant magnetic field amplitude. (b) Microwave transmission $|S_{21}|^2$ as a function of frequency for an input power at the LER of $-115\,\deci\bel\milli$ at a temperature of $20\,\milli\kelvin$, including a fit (solid orange line). (c) Loaded quality factor $\QL$ as a function of the estimated average photon number in the LER at zero magnetic field (purple symbols) and a $1\,\tesla$ in-plane magnetic field (red diamonds). The top axis gives the corresponding microwave power at the resonator. The estimate is a coarse guide, with an uncertainty of one order of magnitude \cite{supplement}. \label{fig:res_overview}}
\end{figure}

Figure \ref{fig:res_overview}\,(a) shows a schematic of the lumped element resonator, which was designed for a high field resilience by minimizing the area of the superconducting thin film. The AC electric and magnetic fields are spatially separated (see Supplementary Information for finite element simulations \onlinecite{supplement}. This allows us to concentrate the magnetic fields around the narrow inductor wire (to strongly couple to a small number of spins), but also introduces significant radiative losses. To suppress the radiative losses, the resonator is placed inside a 3D copper cavity ($Q_{\mathrm{3Dcav}} \approx 800$) and is excited/read-out by capacitively coupling to two antennae protruding inside the 3D cavity volume \cite{Bienfait2016}. Measured in this way, resonators can demonstrate loaded quality factors exceeding $10^5$. 

The resonator shown in Fig. \ref{fig:res_overview}\,(a) has an overall dimension of $600\,\micro\meter \times 600\,\micro\meter$. The capacitor fingers are $10\,\micro\meter$ wide, separated by $50\,\micro\meter$ and the total length of the outer and inner fingers are $1.6\,\milli\meter$ and $1.35\,\milli\meter$, respectively. The inductor wire is $440\,\micro\meter$ long and $2\,\micro\meter$ wide (highlighted yellow in Fig. \ref{fig:res_overview}\,(a)). The resonator is fabricated by electron beam lithography and reactive ion etching into a $\approx 50\,\nano\meter$ thick NbN film, sputtered on a $250\,\micro\meter$ thick high-resistivity ($\rho > 5000\,\ohm\centi\meter$) n-type Si substrate. 
The 3D cavity loaded with the LER is mounted inside a dilution refrigerator and cooled to a base temperature of $20\,\milli\kelvin$. Static magnetic fields of arbitrary orientation were applied using an American Magnetics Inc 3-axis vector magnet (see Supplementary Information for further details on the used measurement setups \onlinecite{supplement}).

Figure \ref{fig:res_overview}\,(b) shows the microwave transmission $|S_{21}|^2$ as a function of frequency at a temperature of $20\,\milli\kelvin$, with an input power at the resonator of $-115\,\deci\bel\milli$ and no externally applied magnetic field. The resonator response is asymmetric due to the strong impedance mismatch induced by the coupling antennae of the 3D cavity \cite{Geerlings2012, Khalil2012}. This can be fit by a Fano resonance \cite{Fano1961} to extract the resonator parameters: frequency $\wc/2\pi = 6375\,\mega\hertz$ and loaded quality factor  $\QL = 2.97\times 10^5$. Figure \ref{fig:res_overview}\,(c) compares $\QL$ as a function of the estimated average photon number $\langle n \rangle$ in the lumped element resonator at zero applied field, versus that at an applied in-plane magnetic field of  $1\,\tesla$. The uncertainty in $\langle n \rangle$ is about an order of magnitude and originates from our estimation of the total attenuation of the setup \cite{supplement}. The zero field loaded quality factor exhibits a kink at $\langle n \rangle \approx 8400$ ($-120\,\deci\bel\milli$) and then continues to increase with increasing microwave power. We attribute this to the onset of nonlinearity, which is accompanied by a downwards shift in frequency (see Supplementary Information \onlinecite{supplement}).
The power dependent data are fit to a two level system (TLS) model, where the quality factor is limited by fluctuating TLSs in the substrate and at the surface \cite{Goetz2016, Burnett2017, Burnett2018} (see Supplementary Information for details on the model \onlinecite{supplement}). The fit is performed for average photon numbers where the resonator is not in the nonlinear regime and is shown by dashed lines in Fig.\,\ref{fig:res_overview}\,(c). This model fits our data well, supporting the interpretation of power dependent losses. Importantly, the loaded quality factor of the resonator remains higher at $1\,\tesla$ than at zero field for all powers where the resonator is in the linear regime. The field dependence of the low-power TLS-limited quality factor suggests that at high field either the TLS states become unpopulated or become detuned from the resonator. However, to fully quantify this observation a more thorough magnetic field dependent study is required, which is beyond the scope of this article, but may be relevant to the impact of TLSs on qubit coherence times \cite{Burnett2019}.
From the measured resonance frequency and an estimate of the LER's capacitance, using conformal mapping techniques \cite{Gevorgian1995}, we determine the resonator's impedance to be $Z = 320\,\ohm \pm 20\,\ohm$.

\begin{figure}[t]
 \includegraphics[]{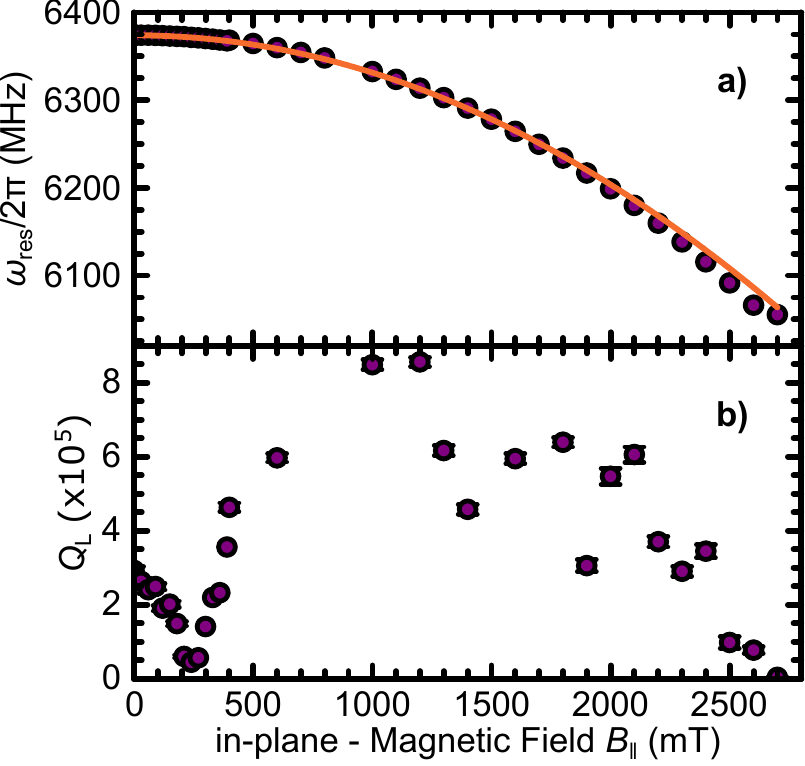}
 \caption{Extracted resonance frequency (a) and loaded quality factor (b) as a function of the in-plane magnetic field $B_{\parallel}$ at a temperature of $20\,\milli\kelvin$ and an input power at the resonator of $-112\,\deci\bel\milli$. The field range from $0 - 400\,\milli\tesla$ is measured with higher resolution. \label{fig:res_ip}}
\end{figure}

Figure \ref{fig:res_overview}\,(a) illustrates the coordinate system we define, in which we create a total magnetic field vector $B_{\mathrm{eff}}$ by applying a constant in-plane field $B_{\parallel}$, together with a smaller perpendicular component $B_{\perp}$ whose angle $\alpha$ is varied.
$B_{\parallel}$ is primarily responsible for setting the overall magnetic field amplitude and direction, which tunes the spin transition frequencies onto resonance with the resonator, and oriented along the inductor so that spins directly beneath the wire satisfy the electron spin resonance condition, whereby the static magnetic field is perpendicular to the oscillating microwave magnetic field.
The orientation for $B_{\parallel}$ is roughly set along a principal axis of the vector magnet when loading the sample, and then carefully aligned to be in the plane of the superconductor through an iterative process at base temperature. We apply a small field ($2\,\milli\tesla$) along the nominal $B_{\parallel}$ axis and then tilt the applied field out of the plane of the superconducting film. At these small fields we can apply the field perpendicular to the resonator without degrading the resonator and thus large tilt angles may be used. By identifying the orientation where the resonator frequency is maximized, we identify an axis which is in the plane of the superconducting thin film. We then ramp the re-defined $B_{\parallel}$ to a larger field, and repeat this process. As the magnitude $B_{\parallel}$ increases, the tilt angle decreases ensuring that large fields are not applied perpendicular to the resonator plane. During this process, we keep the perpendicular field component always smaller than $4\,\milli\tesla$. We choose logarithmically increasing $B_{\parallel}$ setpoints at which we perform the tilting process and complete the alignment with 10 iterations. The duration of the procedure also depends on the magnetic field ramp-rate, which was $50\,\milli\tesla/\mathrm{min}$ and was completed within $\sim\,1.5$ hours.
We followed this alignment process up to $B_{\parallel} = 1\,\tesla$, achieving an accuracy of the in-plane vector of $0.2\,\%$. Although this sets tight bounds on the alignment of $B_{\parallel}$ within the plane of the superconductor, the orientation along the inductor wire was not optimized beyond that upon sample loading. This does not affect the measurements presented here, and alignment could be performed by e.g. maximizing an ESR echo amplitude for spins beneath the wire.

Figure \ref{fig:res_ip} shows the measured resonator frequency and loaded quality factor $\QL$ as a function of the in-plane magnetic field $B_{\parallel}$, while $B_{\perp}$ is kept at $0\,\tesla$.
As the static magnetic field increases from zero to $B_{\parallel} = 2.7\,\tesla$, the resonance frequency decreases by $245\,\mega\hertz$ and largely follows a parabolic dispersion (solid curve), as expected from the kinetic inductance resulting from the change in the Cooper pair density $\ns$ \cite{Wallace1991, Healey2008, Xu2019}. 
The parabolic dispersion only holds for superconductors where vortex losses are not dominant, and a divergence from this behavior indicates that the superconductor is predominately in its type-II state where flux vortices are the main source of loss \cite{Song2009}. For $B_{\parallel} > 2.1\,\tesla$ the resonator frequencies deviate from the parabolic function and for $B_{\parallel} > 2.6\,\tesla$ a kink is observed, which we interpret as that vortex losses become a dominant loss mechanism at such fields. 

As $B_{\parallel}$ is increased from zero, $\QL$ of the resonator drops from about $3 \times 10^5$ to a minimum of about $4 \times 10^4$ at a magnetic field of $234\,\milli\tesla$. We attribute this to the presence of paramagnetic dangling bond defects at the Si/SiO$_2$ (natural oxide) interface, with g-factors $\approx 2$, inductively coupling to the resonator. 
Dangling bond defects \cite{Huebl2008, Pierreux2002, Lenahan1998} are known to have densities of $\approx 10^{12}\,/\centi\meter^2$ and are located in close vicinity to the NbN inductor where the strongest oscillating magnetic fields are present, hence they will strongly interact with the resonator, causing a drop in quality factor due to their dissipation. This is consistent with recent observations on dangling bond defects with g $\approx 2$ reducing the quality factor of resonators on both in silicon \cite{Samkharadze2016} and sapphire \cite{Kroll2019, Graaf2017} substrates at relevant magnetic fields.
Increasing $B_{\parallel}$ further leads to an increase in $\QL$, reaching a maximum of $8.6\times10^5$ at $1\,\tesla$. This suggests that the dangling bond defects limit resonator losses even at zero magnetic field. Note, that the microwave power dependence at $1\,\tesla$ in Fig.\,\ref{fig:res_overview}\,(c) is performed in a different setup where higher field noise limits the maximal achievable $\QL$, resulting in a lower $\QL$ than in Fig.\,\ref{fig:res_ip}\,(b) \onlinecite{supplement}.
For $B_{\parallel} > 1\,\tesla$ the quality factor starts to decrease due to finite misalignments in the static field, as the alignment procedure was performed only up to $B_{\parallel} = 1\,\tesla$. At fields larger than $2.5\,\tesla$ $\QL$ falls below $10^5$. 

\begin{figure}[t]
 \includegraphics[]{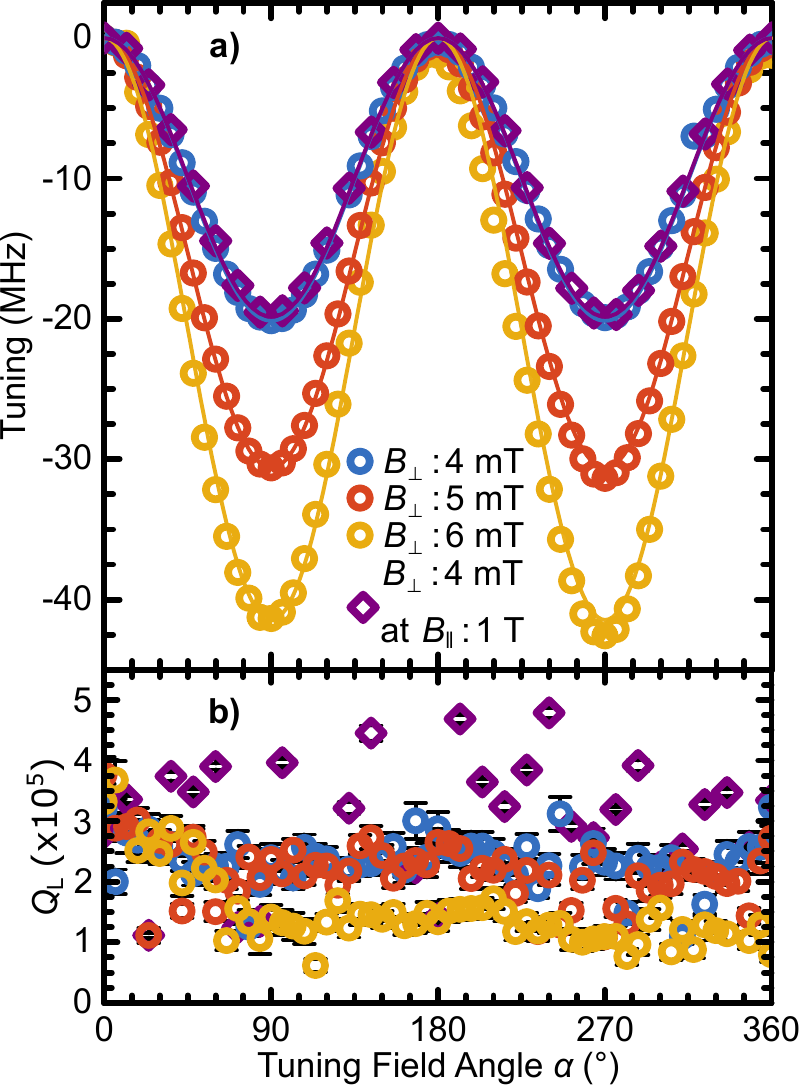}
 \caption{Extracted resonance frequency (a) and loaded quality factor (b) as a function of the out-of-plane magnetic field angle $\alpha$ at a temperature of $20\,\milli\kelvin$ and an input power at the resonator of $-120\,\deci\bel\milli$. The four datasets are rotations with a field magnitude of $B_{\perp}$ of $4\,\milli\tesla$ (blue, circle), $5\,\milli\tesla$ (red, circle), $6\,\milli\tesla$ (yellow, circle) and $B_{\perp}$ of $4\,\milli\tesla$ at $B_{\parallel} = 1\,\tesla$ (dark blue, diamond). The solid lines in (a) showing a calculated $1 + \cos\left(2\alpha\right)$ dependence of the tuning. \label{fig:res_tune}}
\end{figure}

Finally, we investigate the tunability of the resonator frequency by introducing an additional field, $B_{\perp}$, and rotating it by the angle $\alpha$, as shown in Figure \ref{fig:res_overview}\,(a). $B_{\perp}$ is kept smaller than the out-of-plane critical field (estimated to be $B_{\perp,c1} \approx 6.2\,\milli\tesla$) to ensure non-hysteretic frequency tuning. 
Figure \ref{fig:res_tune}\,(a) shows the measured resonator frequency as a function of $\alpha$ for $B_{\perp} = 4 - 6\,\milli\tesla$, at zero applied $B_{\parallel}$, as well as for $B_{\perp} = 4\,\milli\tesla$, with a larger in-plane $B_{\parallel} = 1\,\tesla$.
After each full magnetic field rotation, the resonator is thermally cycled to $18\,\kelvin$ to remove any trapped flux and establish a common reference. This is necessary as although the frequency tuning is non-hysteretic, the resonator loaded quality factor does show hysteresis and does not fully recover to the $0\degree$ value when rotated by $360\degree$, particularly for a $6\,\milli\tesla$ out-of-plane field, as shown in Fig.\,\ref{fig:res_tune}\,(b).
The resonator frequency shows a $1 + \cos\left(2\alpha\right)$ dependence (solid lines in Fig. \ref{fig:res_tune}\,(a)), with a frequency minima for maximal out-of-plane field. The behaviour is symmetric for $B_{\perp} = 4\,\milli\tesla$, while some asymmetry becomes apparent for larger values for $B_{\perp}$ which we attribute to induced flux vortices. 
We define the variability of the resonance frequency tuning as $\sfrac{\wc\left(\alpha\right)}{\wc\left(\alpha+\pi\right)}$, which is below $0.005\,\%$, $0.015\,\%$ and $0.045\,\%$ for $B_{\perp} = 4\,\milli\tesla, 5\,\milli\tesla$ and $6\,\milli\tesla$, respectively.
The maximum tuning range is $20.13(1)\,\mega\hertz$, $30.63(3)\,\mega\hertz$ and $41.4(1)\,\mega\hertz$ for a $B_{\perp}$ of $4\,\milli\tesla$, $5\,\milli\tesla$ and $6\,\milli\tesla$, respectively. At an in-plane field of $1\,\tesla$ the tuning behavior is nearly identical to the zero field case. Here, for $B_{\perp} = 4\,\milli\tesla$ the variability is below $0.005\,\%$ and the maximum tuning range is $19.77 \pm 0.1\,\mega\hertz$, a reduction of tuning range by less than $2\,\%$ compared with the range at zero field.

The loaded quality factor is shown as a function of the magnetic field angle $\alpha$ in Fig. \ref{fig:res_tune}\,(b), and has a value of $3.2 \times 10^5$ for the three different $B_{\perp}$ amplitudes for $\alpha = 0$, with no additional in-plane field. Rotating $B_{\perp}$ out-of-plane of the superconducting film decreases the quality factor: For $4\,\milli\tesla$ and $5\,\milli\tesla$ rotation, $\QL$ drops to an average value of $2.2 \times 10^5$ when $\alpha$ reaches $90\,\degree$ and remains constant for the rest of the rotation. The drop for $B_{\perp} = 6\,\milli\tesla$ rotation is more significant, falling to a value of $1.3\times10^5$ then again remaining constant. The initial drop in $\QL$ indicates the generation of flux vortices even at small perpendicular magnetic fields, however for these values of $B_{\perp}$ the losses are tolerable as a $\QL > 10^5$ can be maintained and no hysteretic behavior in resonance frequency is observed for $B_{\perp} = 4\,\milli\tesla$ and only a small hysteretic effect for the higher perpendicular fields.
At $B_{\parallel} = 1\,\tesla$ and $B_{\perp} = 4\,\milli\tesla$, $\QL$ maintains an average value of about $4\times10^5$. At these static in-plane magnetic fields, noise from the magnet is believed to limit the stability of the LER's resonance frequency, leading to a scatter in the measured $\QL$.

Although the primary motivation of the methods presented here is the relatively slow tuning of the resonator frequency to match a desired spin transition, it is also worth reflecting on potential applications in fast-tuning of the resonator frequency within a quantum memory pulse sequence \cite{Kubo2011}. Tuning the resonator frequency by one resonator linewidth ($\approx 28\,\kilo\hertz$) would require $B_{\perp} \approx 140\,\micro\tesla$ and given the maximum magnetic field ramp-rate ($200\,\milli\tesla/\min$) of the magnet systems used, this could be achieved within $42\,\milli\second$. Low inductance magnetic coils such as modulation coils used in conventional ESR \cite{Poole1997} can apply magnetic fields of $\sim 1\,\milli\tesla$ at a frequency of $100\,\kilo\hertz$ i.e. a $10\,\micro\second$ field tuning. This is considerably faster than the coherence time for spins at these magnetic field - frequency optimal working points and could therefore be used to tune resonators within pulse sequences for quantum memory experiments.

In summary, we presented a design for a high-quality factor, co-planar superconducting lumped element microwave resonator made of NbN, which can be operated at high static magnetic fields (up to $2.6\,\tesla$ in-plane of the superconductor), while maintaining a high-quality factor ($>10^5$). We observe a significant drop in quality factor arising from coupling to $g \approx 2$ spins, most likely dangling bond defects at the Si/SiO$_2$ interface.
We demonstrated the tuning of the resonator frequency by applying a small magnetic field perpendicular to the superconducting film, and we see near non-hysteretic frequency tuning up to $30.63(3)\,\mega\hertz$, while maintaining the high-quality factor. The tuning range can be further increased with higher perpendicular fields, however, the resonance frequency tuning becomes hysteretic and the quality factor drops. Similar tuning can be performed using significant in-plane fields (e.g.\ $1\,\tesla$). This type of resonator is therefore well suited to study the spin-resonator coupling at specific combinations of magnetic field magnitudes and resonance frequencies, e.g. magnetic field noise resilient transitions, and has a high potential for devices such as quantum memories.

See the Supplementary Information for a detailed description of the modelling of the resonator power dependence, the resonator fabrication finite element simulations, the experimental setups, the reproducibility of the resonator parameters between cooldowns and a characterization of additional device.

This research has received funding from the European Union's Horizon 2020 research and innovation programme under grant agreement No 688539 (\href{http://mos-quito.eu}{MOS-QUITO}), as well as from UK Engineering and Physical Sciences Research Council, Grant No. EP/P510270/1.

\pagebreak
\widetext
\begin{center}
\textbf{Tuning high-Q superconducting resonators by magnetic field reorientation: Supplemental Information}
\end{center}

\setcounter{equation}{0}
\setcounter{figure}{0}
\setcounter{table}{0}
\setcounter{page}{1}
\makeatletter
\renewcommand{\thefigure}{S\arabic{figure}}
\renewcommand{\theequation}{S\arabic{equation}}
\renewcommand{\thetable}{S\arabic{table}}
\renewcommand{\bibnumfmt}[1]{[S#1]}
\renewcommand{\citenumfont}[1]{S#1}
\newcommand{\hbl}[1]{\textcolor{red}{#1}}
\newcommand{\cwz}[1]{\textcolor{blue}{#1}}

\setlength{\paperheight}{297mm}
\setlength{\paperwidth}{210mm}

\section{Modelling the Microwave Power Dependence of the Lumped element resonator}

Here, we discuss the microwave power dependence of the loaded quality factor $\QL$. In Fig. 1\,(c) of the main text, we fit a two level system (TLS) model to the measured values of $\QL$ as a function of input power, for both zero applied field and at $1\,\tesla$. These models describe the loss tangent $\delta$ of a resonator which is defined by the internal Q-factor $\delta = \sfrac{1}{Q_{\mathrm{i}}}$. For the experiments presented in the main text, the superconducting resonator is placed inside a 3D copper cavity to suppress radiative losses. The superconducting resonator is only coupled weakly via two antennas, protruding into the 3D cavity volume, which results in very high external quality factors ($Q_{\mathrm{e}} \gg 10^6$). The total loaded Q-factor is given by the reciprocal sum of the internal and external Q-factors $\sfrac{1}{\QL} = \sfrac{1}{Q_{\mathrm{e}}} + \sfrac{1}{Q_{\mathrm{i}}}$. As ${Q_{\mathrm{e}}} \gg {Q_{\mathrm{i}}}$, we can assume that $\QL \approx {Q_{\mathrm{i}}}$ and thus $\delta = \sfrac{1}{\QL}$. 
We use the following model \cite{S_Burnett2017}:
\begin{equation}
 \delta\left( P_{\mathrm{in}}, T \right) = \delta_{\mathrm{TLS,0}} \frac{\tanh\left( \sfrac{\hbar \wc}{2k_{\mathrm{B}}T} \right)}{\left( 1 + \sfrac{\langle n \rangle}{n_{\mathrm{c}}} \right)^{\beta}} + \delta_{\mathrm{other}},
 \label{eqn:TLS_model}
\end{equation}
where $\delta_{\mathrm{TLS,0}}$ represents the loss tangent for the unsaturated TLS in the low temperature and low power limit, $\wc / 2\pi$ the resonator resonance frequency, $k_{\mathrm{B}}$ the Boltzman constant, $\langle n \rangle$ the average number of microwave photons in the resonator, $n_{\mathrm{c}}$ the critical photon number for saturating the TLS, 
$\beta$ the rate of saturation of the TLS, and $\delta_{\mathrm{other}}$ comprises the loss tangent of other power independent losses, like quasiparticle losses, radiative losses etc. \cite{S_Goetz2016}. The average number of photons in the resonator is defined by the input microwave power at the resonator $P_{\mathrm{in}}$, its resonance frequency and full width at half maximum $\Delta f$, $\langle n \rangle = \sfrac{10^{-3} \times 10^{P_{\mathrm{in}}/10}}{\hbar \wc \Delta f}$. The results of fitting \eqref{eqn:TLS_model} to the datasets in Fig.\ 1\,(c) of the main text are summarized in Tab.\,\ref{tab:TLS-paras}. The large error is due to the small number of data points, which increases the uncertainty of the fitted parameters. Nevertheless, the data can be described by a well-known model, and the presented resonator design behaves within its expectations. 

\begin{table}[b]
\center
 \begin{tabular}{|c|c|c|c|c|}
  \hline
  $B_{\parallel}$ $(\tesla)$ & $n_{\mathrm{c}}$ & $\beta$ & $\delta_{\mathrm{TLS,0}}$ $(\times 10^{-4})$ & $\delta_{\mathrm{other}}$ $(\times 10^{-6})$ \\ \hline
  0 & $7.80 \pm 12.35$ & $0.47 \pm 0.15$ & $5.29 \pm 2.10$ & $4.12 \pm 2.63$ \\ \hline
  1 & $5.74 \pm 12.21$ & $0.47$ & $1.58 \pm 1.23$ & $2.57 \pm 0.09$ \\ \hline
 \end{tabular}
 \caption{Summary of the extracted parameters, using \eqref{eqn:TLS_model}. We used the zero field $\beta$ fit result to fit the $1\,\tesla$ dataset.}
 \label{tab:TLS-paras}
\end{table}

At low microwave powers the unsaturated TLSs $\delta_{\mathrm{TLS,0}}$ are limiting the quality, which are coupling to the electromagnetic field of the resonator. Typically, the loaded quality factor increases with microwave input power until it becomes limited by power independent losses $\delta_{\mathrm{other}}$. Interestingly, the $\QL$ remains higher at $1\,\tesla$ than at zero field for all $\langle n \rangle$ where the resonator is in the linear regime. The field dependence of the low-power TLS-limited quality factor suggests that at high field either the TLS states become unpopulated or become detuned from the resonator.

\section{Nonlinear Microwave Power Regime}

\begin{figure}[b]
 \includegraphics[]{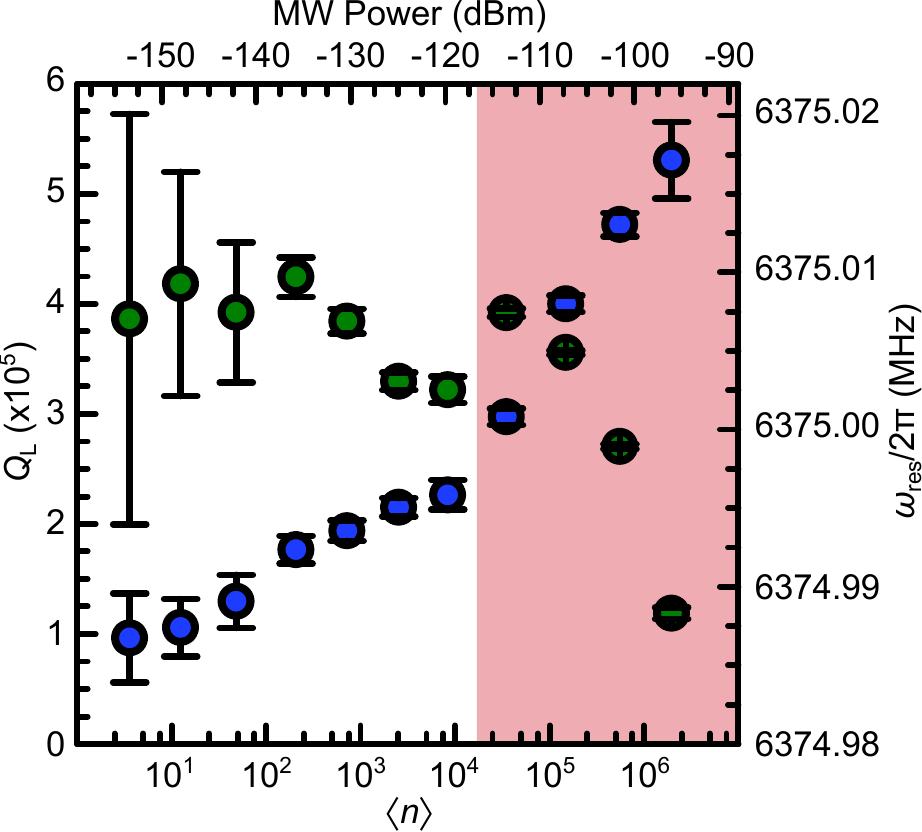}
 \caption{Loaded quality factor $\QL$ (left axis, blue symbols) and resonance frequency $\wc / 2\pi$ (right axis, green symbols) as a function of average photon number $\langle n \rangle$ in the resonator. The top axis shows the respective microwave input power at the resonator. The red underlaid area indicates the nonlinear regime. \label{S1}}
 \end{figure}
Figure \ref{S1} shows the zero magnetic field loaded quality factor $\QL$ (left axis, blue symbols) as a function of the average number of photons in the resonator $\langle n \rangle$ (cf. Fig. 1\,(c) in the main text). In addition, the resonance frequency $\wc / 2\pi$ is plotted (right axis, green symbols) as a function of $\langle n \rangle$. We observe a kink at a photon number of about $10^4$, corresponding to an input power of $-120\,\deci\bel\milli$. We attribute this kink to the onset of the nonlinear regime of the resonator. The nonlinearity arises from the kinetic inductance of the superconductor and results in a Duffing oscillator behaviour \cite{S_Zmuidzinas2012, S_Burnett2017}. A clear signature of this regime is a strong asymmetry in the resonance lineshape, which we observe at the highest powers applied, indicating that at these points the resonator is well in the nonlinear regime. This effect is accompanied by a resonance frequency shift, due to the nonlinear kinetic inductance. As shown in Fig.\,\ref{S1} the LER's resonance frequency is shifted downwards for $\langle n \rangle > 10^4$ or corresponding $P_{\mathrm{in}} > -120\,\deci\bel\milli$. We take this point, where the resonator frequency begins to shift downwards, as the onset of the nonlinearity.

\section{Resonator Fabrication and Simulations}

Our resonators are fabricated by electron beam lithography (Raith 150-TWO) and reactive ion etching (Oxford Plasma Pro NGP80) into a $\approx 50\,\nano\meter$ thick NbN film, sputtered on a $250\,\micro\meter$ thick high-resistivity ($\rho > 5000\,\ohm\centi\meter$) n-type Si substrate. 
Prior to the metal deposition the natural oxide layer on the Si substrate is removed by a $10\,\sec$ HF dip and directly transferred into the sputter chamber (less than $1\,\min$). The NbN film was sputtered in a SVS6000 chamber, at a base pressure of $6.5 \times 10^{-7}\,\milli\bbar$, using a sputter power of $200\,\watt$ in an 50:50 Ar/N atmosphere held at $5 \times 10^{-3}\,\milli\bbar$, with the gas flow for both elements set to $50\,$SCCM. The resulting NbN films showed a critical temperature of $T_{\mathrm{c}} = 11.6\,\kelvin$. Using the same deposition conditions $20\,\nano\meter$ thick NbN films were deposited on sapphire and have been previously demonstrated to have internal quality factors at high power (low power) of over $10^6$ ($4\times 10^5$) in a quarter wave coplanar waveguide resonator, with $T_{\mathrm{c}} \sim 10\,\kelvin$ \onlinecite{S_Burnett2017}. The difference in $T_{\mathrm{c}}$ between that work and the devices studied here are ascribe to film thickness.

\begin{figure}[t]
 \includegraphics[]{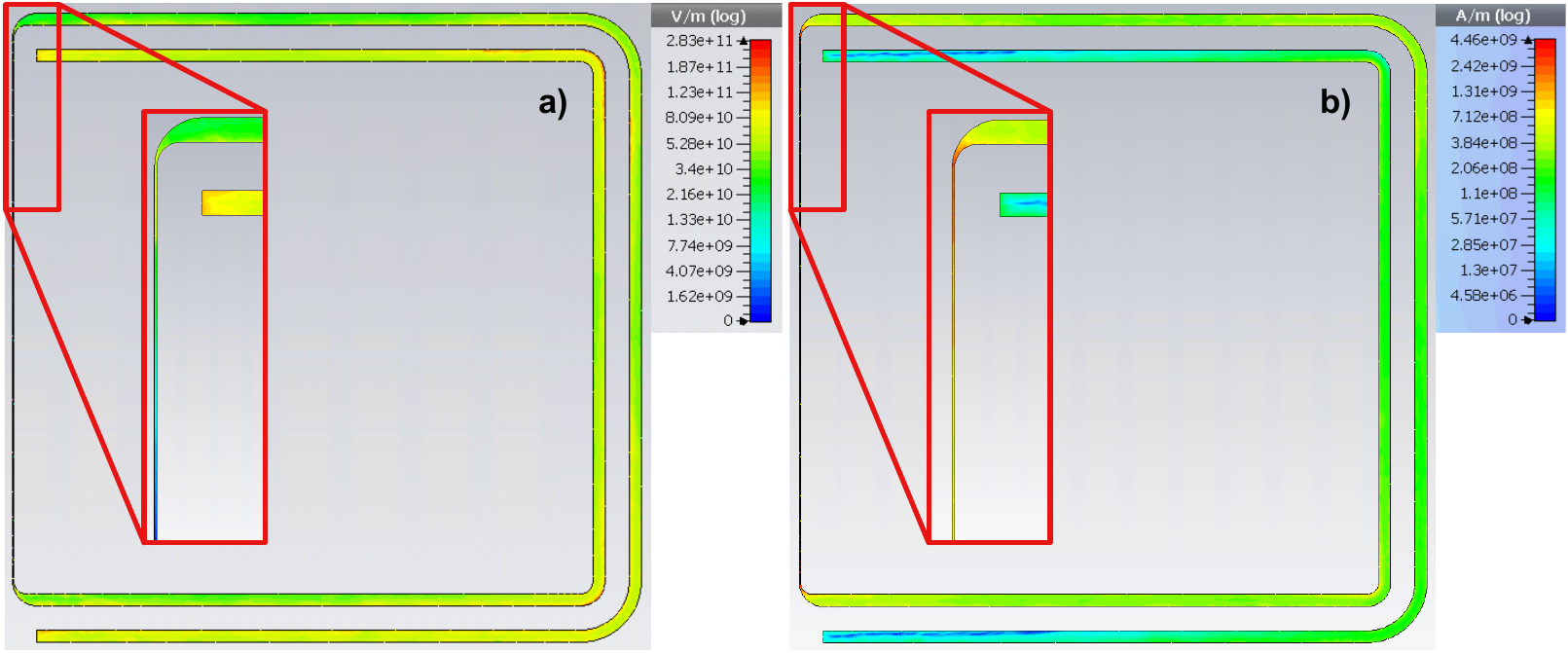}
 \caption{CST Microwave Studio simulations for the superconducting resonator design, showing the distribution of $E$-fields (a) and $H$-fields (b) of the resonant mode. \label{S2}}
 \end{figure}

We use the finite element simulation software CST Microwave Studio to simulate the superconducting resonator design, presented in the main text. Using the eigenmode solver, we plot the distributions of the $E$-fields and the $H$-fields of the resonant mode in Fig.\,\ref{S2}\,(a) and (b), respectively. The $H$-field density is maximal within the narrow inductive wire segment of the resonator and more than an order of magnitude lower in the capacitive part of the design. Opposite to this, is the distribution of $E$-fields, which approaches zero in the inductive part and maximal in the capacitive part. The $E$-fields and $H$-fields are well separated in this design, making it well suited for spin resonance experiments.

\section{Measurement Setup}\label{sec:setup}

The experiments presented in the main text were performed in two different dilution refrigerators. Both are of the same model (BlueFors LD400), but they differ in their American Magnetics Inc 3D vector magnet system and their total microwave circuit attenuation. The magnet system used for the tuning experiment (Fig.\,3 in the main text) and the microwave power sweep at $B_{\parallel} = 1\,\tesla$ (Fig.\,1\,(c) in the main text) is a three split-coil superconducting 3D vector magnet system ($3-1-1\,\tesla$), without superconducting persistent switches. With this magnet system we observed increased magnetic noise at static magnetic field vectors $> 500\,\milli\tesla$. The in-plane field resilience experiment (Fig.\,2 in te main text) and the zero field microwave power sweep (Fig.\,1\,(c) in the main text) are performed with a solenoid and two split-coils superconducting 3D vector magnet system ($6-1-1\,\tesla$), with superconducting persistent switches. This system showed more stable fields and less magnetic field noise, enabling magnetic fields of up to $2.6\,\tesla$ to be applied while maintaining a stable resonance frequency of the superconducting LER.
The decrease and scatter in the measured $\QL$ may arise from larger magnetic field noise in the ($3-1-1\,\tesla$) system. The high Q-factor together with the large kinetic inductance of the presented resonator design makes it especially sensitive to magnetic field fluctuations. Small fields of the order of $\mu\tesla$ are enough to shift the resonator frequency by a linewidth and thus influence the measured $\QL$. 

\begin{figure}[ht]
 \includegraphics[]{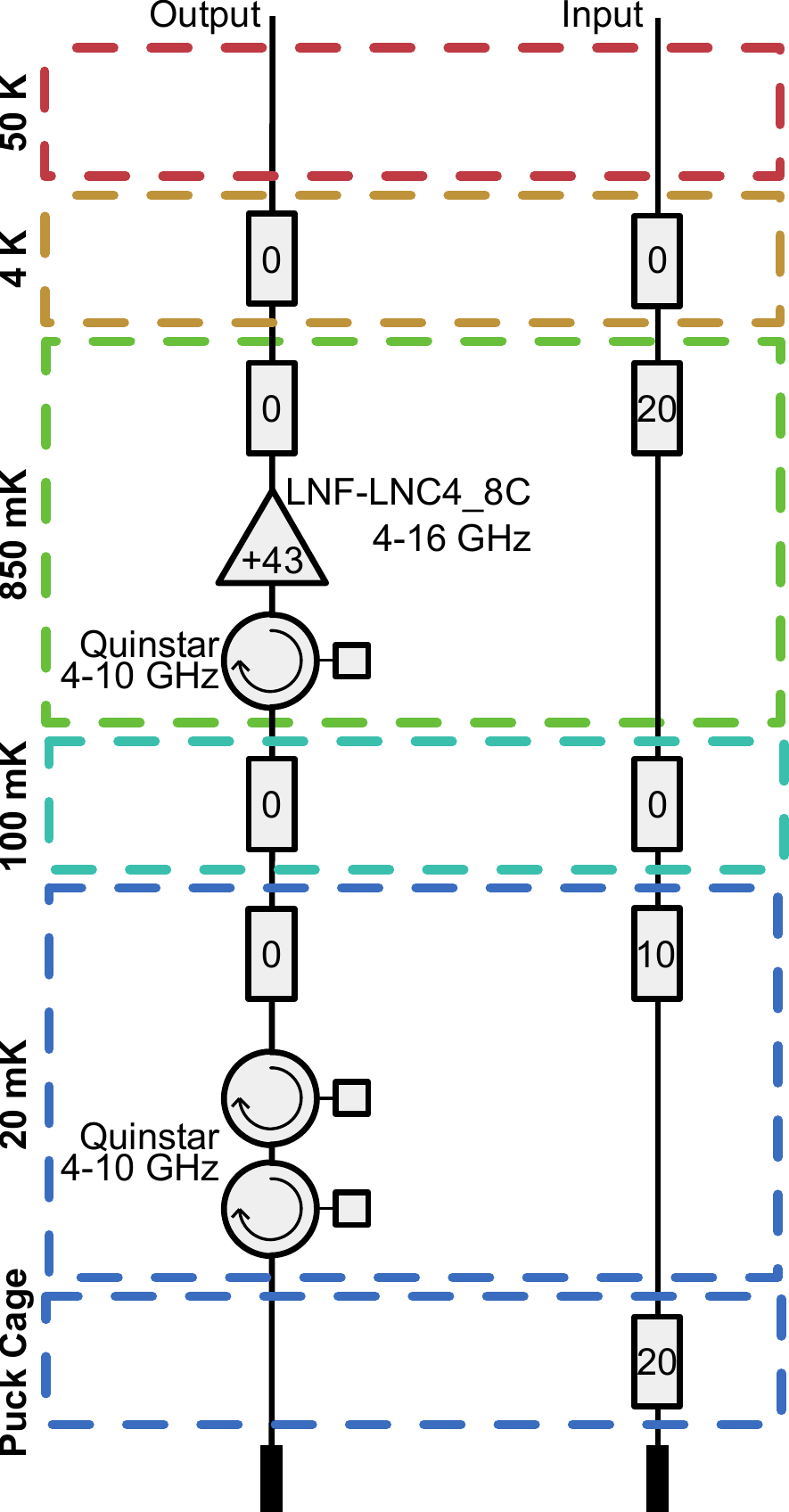}
 \caption{Schematic of the microwave circuitry mounted in the two dilution refrigerators used. The different temperature stages are indicated by the dashed rectangular boxes and the components mounted on these stages. \label{S3}}
 \end{figure}

Figure \ref{S3} shows a schematic of the microwave circuitry used in our experiments. The microwave input line in the cryostat is attenuated by $50\,\deci\bel$ between room temperature and the mixing chamber stage to reduce thermal noise and thermally anchor the center conductor of the coaxial cables, while the output line contains three cryogenic isolators, two at base temperature and one at still temperature ($\approx 850\,\milli\kelvin$), to suppress thermal noise reaching the sample. The output signal is amplified at $4\,\kelvin$ by a cryogenic HEMT amplifier ($+40\,\deci\bel$) and then at room temperature ($+20\,\deci\bel$). The microwave components are similar for both dilution refrigerators. The setup differs in the cabling outside the cryostat. The tuning experiment (Fig.\,3 in the main text) and the microwave power sweep at $B_{\parallel} = 1\,\tesla$ (Fig.\,1\,(c) in the main text) are using long microwave coaxial cables to connect to the vector network analyzer ($\approx 10\,\meter$ in total). The cables used for the in-plane field resilience experiment (Fig.\,2 in te main text) and the zero field microwave power sweep (Fig.\,1\,(c) in te main text) are considerably shorter ($\approx 10\,\meter$ in total). This results in a difference in total attenuation for both systems of about $10\,\deci\bel$. The difference has been accounted for in the main text. Further, we estimate the total attenuation of the two setups to $100\,\deci\bel$ and $90\,\deci\bel$, respectively. It includes the microwave circuitry ($\approx 75\,\deci\bel$ and $\approx 65\,\deci\bel$, respectively) and the insertion loss into the 3D cavity and to the superconducting LER ($\approx 25\,\deci\bel$). The  uncertainty in our estimate lies mostly in the insertion loss to the lumped element resonator, due to the 3D cavity readout scheme we use in our experiments.

\section{Reproducibility Between Cooldowns}

\begin{figure}[ht]
 \includegraphics[]{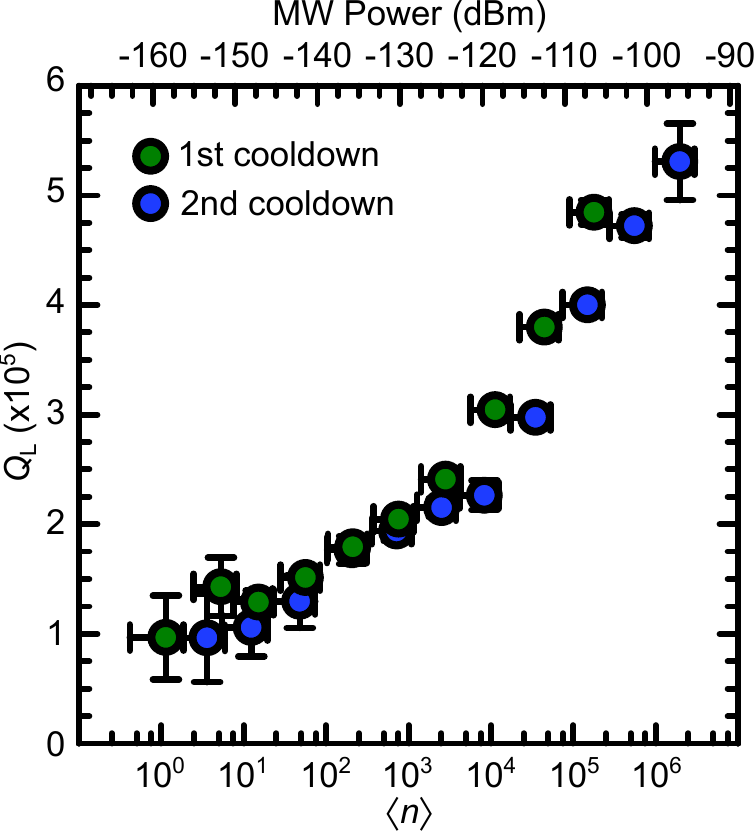}
 \caption{Loaded quality factor $\QL$ as a function of the estimated average photon number at $20\,\milli\kelvin$ and for zero applied external magnetic field. The top axis shows the estimated input microwave power at the resonator. The two datasets are from two separated cooldowns. \label{S4}}
 \end{figure}

The superconducting lumped element resonator, presented in the main text, is characterised in two different dilution refrigerators (see section \ref{sec:setup}). This allows us to compare the resonator parameters from two distinct cooldowns and quantify the device's reproducibility. Figure \ref{S4} compares the loaded quality factor $\QL$ as a function of the estimated average photon number in the resonator for two separate cooldowns, at $20\,\milli\kelvin$ and zero applied external magnetic field. The first cooldown is performed in the dilution refrigerator with the $3-1-1\,\tesla$ magnet system and the second in the dilution refrigerator with the $6-1-1\,\tesla$ magnet system (c.f. section \ref{sec:setup}). $\QL$ for both cooldowns coincide and only show a small deviation in the high-power regime ($\langle n \rangle > 10^4$). Between the two cooldowns the sample was stored at atmosphere for four days. The resonance frequencies $\wc/2\pi$ are $6374.35\,\mega\hertz$ and $6375\,\mega\hertz$ for the first and second cooldown, respectively, which corresponds to a change of $0.01\,\%$ between the two runs.
We attribute the high reproducibility to the used superconducting material, NbN. Due to its high nitrogen content, NbN is resilient to oxidation and thus degradation of the resonator's performance even under prolonged exposure to air.

\section{Additional Device Characterisation}

\begin{figure}[t]
 \includegraphics[]{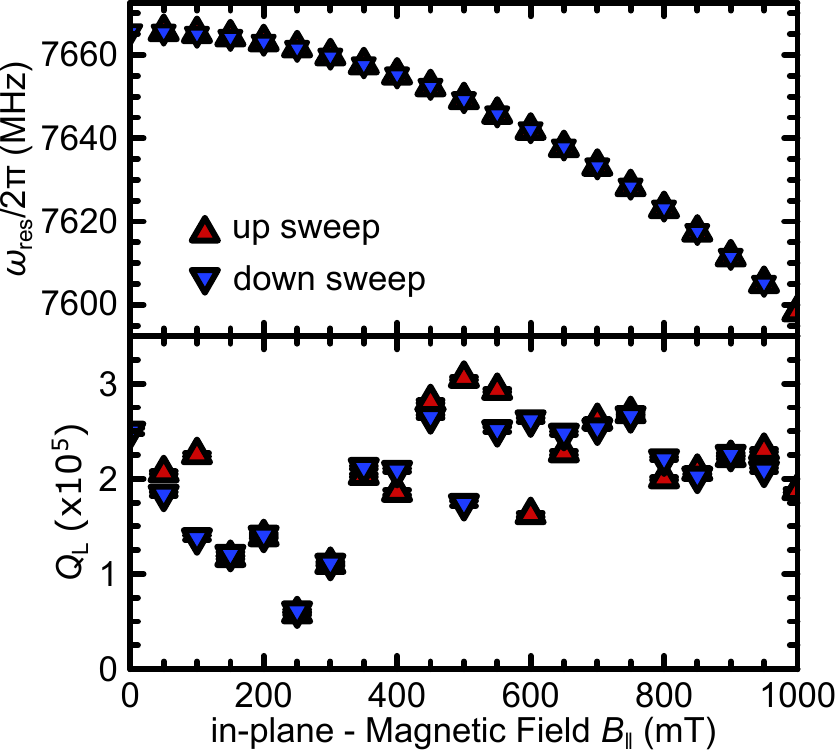}
 \caption{Extracted resonance frequency (a) and loaded quality factor (b) as a function of the in-plane magnetic field $B_{\parallel}$ at a temperature of $20\,\milli\kelvin$ and an input power at the resonator of $-120\,\deci\bel\milli$. The magnetic field is swept up (upwards red triangles) and down (dwonwards blue triangles). \label{S5}}
 \end{figure}

Here, we present additional data acquired on a similar device as presented in the main text. Figure \ref{S5}\,(a) shows the fitted resonance frequency of this resonator as a function of the in-plane magnetic field $B_{\parallel}$. The magnetic field is swept up to $1\,\tesla$ (red triangles) and back to zero (blue triangles). The dependence of $\wc$ is similar to the resonator presented in the main text. Additionally, it shows that there is no visible hysteresis between the up and down sweep. Figure \ref{S5}\,(b) shows loaded quality factor $\QL$ as a function of $B_{\parallel}$. Similar to our device in the main text, $\QL$ shows a minimum at a frequency and magnetic field corresponding to an interaction with $g \approx 2$ spins. For higher magnetic fields $\QL$ increases again until it reduces monotonically with increasing fields. The data of the magnetic field up and down sweep  follow the same dependence, showing the high reproducibility of our resonator design.

\end{document}